\documentclass[journal]{ieeetran}
\usepackage{amsmath,amsthm,amssymb,paralist,subfigure,cite,graphicx}
\usepackage{algorithm2e}
\usepackage[table]{xcolor}

\newtheorem{theorem}{Theorem}

\newtheorem{prob}{Problem Formulation}

\def\mb{\mathbf}
\def\mbb{\mathbb}
\def\mc{\mathcal}

\DeclareMathOperator*{\argmin}{argmin}

\begin{document}
\title{ Structural cost-optimal design of sensor networks for distributed estimation}
\author{Mohammadreza Doostmohammadian$^\dagger$$^\ast$, \textit{Member, IEEE}, Hamid R. Rabiee$^\dagger$, \textit{Senior Member, IEEE},  Usman A. Khan$^\S$, \textit{Senior Member, IEEE}

\thanks{
$^\dagger$ ICT Innovation Center for Advanced Information and Communication Technology, Sharif University of Technology, Tehran, Iran {
\texttt{rabiee@sharif.edu}}.

$^\ast$ (corresponding author) Mechanical Engineering Department, Semnan University, Semnan, Iran \texttt{doost@profs.semnan.ac.ir}.

$^\S$ Electrical and Computer Engineering Department, Tufts University, Medford, USA \texttt{khan@ece.tufts.edu}.}}
\maketitle

\begin{abstract}
	In this letter we discuss cost optimization of sensor networks monitoring structurally full-rank systems under distributed observability constraint. Using structured systems theory, the problem is relaxed into two subproblems: (i) sensing cost optimization and (ii) networking cost optimization. Both problems are reformulated as combinatorial optimization problems. The sensing cost optimization is shown to have a polynomial order solution. The networking cost optimization is shown to be NP-hard in general, but has a polynomial order solution under specific conditions. A 2-approximation polynomial order relaxation is provided for general networking cost optimization, which is applicable in large-scale system monitoring.
	
\end{abstract}

\section{Introduction} \label{sec_intro}
Single time-scale  distributed estimation among a group of sensors/agents
has been the topic of interest in the signal-processing  literature \cite{das2016consensus,mohammadi2015distributed,das2015distributed,sayedtu12,sayed11,kar2013consensus,jstsp,jstsp14, battistelli_cdc,nuno-suff.ness,doostmohammadian2017recovery}.
In this paper, it is assumed that the underlying system is structurally full-rank (also known as structurally cyclic systems), which is also the case in \cite{pequito2017structurally,pequito2013optimal}. 
This is a typical assumption in distributed estimation literature as in \cite{sayedtu12,sayed11,nuno-suff.ness,battistelli_cdc}, where proper sensor measurements (sensing) satisfying observability constraints  along with sufficient information sharing among sensors (networking) provide distributed estimation of system with bounded error. The idea in this paper is to minimize the cost of sensing and networking while satisfying distributed observability constraints. The related literature on this problem (with observability constraint) is limited  \cite{pequito_gsip,pequito2017structurally,pequito2013optimal}  and the problem is to great extent unexplored. To solve the problem, structural relaxations are applied to formulate the problem as proper combinatorial optimization format, knowing that structural relaxations are valid for almost all possible numerical values of system parameters \cite{woude:03}.

In this paper, optimization of sensing and networking cost are considered. Note that state measurements by sensors have certain cost. The sensing cost is similar to cost of sensor selection, and the networking cost is similar to link/communication cost. The sensor measurement cost might be due to, for example, sensor range/calibration, sensor's energy/power consumption, maintenance/embedding expenses for sensor placement, and even environmental condition such as humidity/temperature. On the other hand, every link in the sensor network has a cost, representing, for example, data transmission energy, and communication cost of sensors that may be subject to environmental conditions.
As the first contribution, we show that  the sensing cost optimization problem for structurally full-rank systems has a polynomial order solution with complexity $\mc{O}(m^3)$. This is a significant result, as in \cite{pequito_gsip} authors claim that for general systems there is no polynomial-order solution for sensing cost optimization, i.e. the problem is NP-hard\footnote{NP-hard problems are believed to have no solution in time-complexity upper-bounded by a polynomial function of the input parameters.}. Another related literature,  \cite{boyd2009sensor} considers sensor selection for reducing the measurement error, while in \cite{krause2006near} near-optimal sensor placement based on greedy algorithm is adopted. None of these works consider observability constraint which makes the problem NP-hard in general, and this is the main distinction of our work. Next, we extend to the networking cost optimization. The introductory results on network design based on distributed estimation and formulation of distributed observability are taken from \cite{asilomar11, jstsp}. As the next contribution, this paper generalizes the cost-optimal design of sensor network for \textit{centralized} estimation\cite{pequito2017structurally,pequito2013optimal,pequito2014optimal,pequito2014design} to distributed case while the problem is constraint with \textit{distributed} observability of system/network. Our results extends the leader selection scenario in \cite{pequito2015distributed} by cosidering cost and distributed observability constraints. We show that with bidirectional communications among sensors the networking cost optimization has a polynomial order solution. The most general case, where the communication links are directional, is shown to be NP-hard. For this case, a 2-approximation\footnote{An algorithm is $\rho$-approximation algorithm if it finds a solution within a factor $\rho$ of optimum solution.} algorithm with polynomial complexity $\mc{O}(m^2)$ is suggested.

\section{Problem Formulation} \label{sec_prob}
Consider estimation of LTI system with measurements:
\begin{eqnarray}\label{eq_sys1}
\mb{x}_{k+1} &=& A\mb{x}_k + \mb{v}_k,
\\\label{eq_sys2}
y_k^j &=& H_j\mb{x}_k + r_k^j,\qquad j\in \{1,...,m\}.
\end{eqnarray}
where~$\mb{x}=[x_{1}~\ldots~x_{n}]^\top\in\mbb{R}^n$ is state-space,~$\mb{y}=[y_1,\ldots,y_m]\in\mbb{R}^m$ is measurement vector, $\mb{v}$ and~$\mb{r}$, are noise variables with standard assumptions on Gaussianity and independence. Define $H_j$, the $j$-th row of $H$ with dimention $1$-by-$n$, as the measurement (row) matrix of sensor $j$, and $y_k^j$, the $j$-th column of $\mb{y}_k$, as measurement of sensor $j$. Throughout the paper we may omit the time index~$k$ and use $\mb{y}$. Based on Kalman \cite{kalman:61}, bounded estimation error requires $(A,H)$-observability. The system is monitored by a network of sensors, denoted by $\mc{G}_W$. This paper considers single time-scale distributed estimation over the sensor network, where both system dynamics and distributed estimator evolve at the same time-scale \cite{das2016consensus,das2015distributed,sayedtu12,sayed11,kar2013consensus,jstsp,jstsp14, sauter:09,nuno-suff.ness}. 
In the single time-scale distributed estimation method two types of information-fusion are performed: (i) prediction-fusion, and (ii) measurement fusion. It is known that if the system is structurally cyclic (assumed in this paper), only prediction fusion is necessary for monitoring the global state of the system, see \cite{jstsp,jstsp14,globalsip14} for details.
Every sensor shares the  state prediction of the system over the neighborhood, $\mc{N}$, over the communication network $\mc{G}_W$. Denoted by $W$ the network adjacency matrix, where $W_{ij}$ defines the  consensus weight for prediction fusion. Note that the entries in the adjacency matrix are defined such that $W$ is row-stochastic, see \cite{jstsp,jstsp14,nuno-suff.ness} for details.
Note that, in distributed estimation the global state of the system is observable to every sensor via information-fusion over the sensor network; this is called distributed observability and implies that there exist a feedback gain matrix such that every sensor achieves asymptotic omniscience on the global state of the system and the error dynamics of sensors achieves global asymptotic stability on Mean Squared Error (MSE) \cite{jstsp,jstsp14,nuno-suff.ness}. This is formally stated in the following theorem:

\begin{theorem} \label{thm_stability}
	Given a structurally full-rank system matrix $A$, measurement matrices $H_i, i\in\{1,...,m\}$, communication network of sensors $\mc{G}_W$, if $(W\otimes A, D_H)$ is observable the system is distributed observable by the group of sensors, where $\otimes$ is the kronecker product and $D_H$ is defined as follows:
	\begin{eqnarray}\nonumber
	D_H = blockdiag \left(  H_1^\top H_1, \ldots , H_m^\top H_m\right)
	\end{eqnarray}
\end{theorem}
Recall that distributed observability implies that the system is observable at each sensor. 
The general theorem with detailed proof (irrespective of system structural rank) is given in \cite{jstsp,jstsp14,globalsip14}. The proof for structurally full-rank systems is also provided in \cite{nuno-suff.ness, sayed11}. Note that this result is irrespective of specific algorithms for distributed estimation and holds in general. In other words, structural observability of $(W\otimes A, D_H)$ is sufficient for existance of certain gain matrix for  distributed estimation protocol. Note that in this paper this condition is structural-based, and the exact numerical values can be determined based on specific esimation protocol.

The problem is to determine the structure of measurement/sensing matrix $H$ and the network adjacency matrix $W$ such that $(W \otimes A,D_H)$-observability is satisfied. Note that, as mentioned in the introduction, every choice of sensor measurement and communication network accompanies with a cost.
 On the other hand, all possible communication links among sensors are represented by the network $\mc{G}_{net}$. As mentioned in the introduction, every link in $\mc{G}_{net}$ has a cost, referred to as networking cost. The problem is to minimize both measurement/sensing cost and networking cost (from possible communication links $\mc{G}_{net}$) while satisfying distributed observability of sensors. The problem is formally described below:

\begin{prob} \label{prob_1}
	For a group of sensors assume sensing cost $c_{ij}$ for every sensor $y_i , i \in \{1, \hdots, m\}$ measuring state $x_j , j \in \{1, \hdots, n\}$, and networking cost $b_{ij}$ for communication from sensor $y_i$ to $y_j$ in $\mc{G}_{net}$. Given the network $\mc{G}_{net}$ and the cost matrices $c$ and $b$ the problem is to minimize sensing and networking cost of monitoring the global state of the dynamical system \eqref{eq_sys1}, leading to the following formulation:
	\begin{equation} \label{eq_formulation1}
	\begin{aligned}
	\displaystyle
	\argmin
	\limits_{\mc{H},\mc{W}} ~~ & \sum_{i=1}^{m} \sum_{j=1}^{n} c_{ij}\mc{H}_{ij} + \sum_{i=1}^{m} \sum_{j=1}^{m} b_{ij}\mc{W}_{ij} \\
	\text{s.t.} ~~ & (W \otimes A,D_H)\mbox{-observability},\\
	~~ & \mc{G}_W \subset \mc{G}_{net},\\
	~~ &  \mc{H}_{ij} \in \{0,1\}, ~~\mc{W}_{ij} \in \{0,1\}\\
	~~ &  A~\mbox{is structurally full rank}\\
	\end{aligned}
	\end{equation}
	where $A$ is system matrix, $H$ is measurement matrix, $W$ is adjacency matrix of sensor network, $\mc{H}$ represents the $0\mbox{-}1$ structure of $H$, and $\mc{W}$ represents the $0\mbox{-}1$ structure of $W$.
\end{prob}
The following assumptions are considered:
\begin{itemize}
\item Every sensor is assigned with only one measurement (without loss of generality), and the number of sensors is sufficient for system observability. Note that we assume minimum number of sensors for observability; each sensor measures at least one state and there is no inactivated sensor with no measurements.

\item The given network $\mc{G}_{net}$ representing all possible communications among sensors is Strongly Connected (SC).

\item The optimization problem is solved, not by a leader node, but by user once and then the sensor network is designed.  

\end{itemize}

\section{Structural Relaxation} \label{sec-str}
In this section we provide some graph-theoretic concepts to relax the observability constraint in Problem Formulation~\ref{prob_1}. Reformulating the problem as known combinatorial problems, the solution is provided in the next section. As mentioned in the previous section, observability is required for estimation of dynamical systems.
This paper adopts a graph-theoretic observability method, called structural observability. It is known that such methodology is irrespective
of numerical values of system parameters  and only deals with  system digraph representing the zero-nonzero pattern of system matrix \cite{asilomar11,liu-pnas,woude:03}. Based on this methodology we relax the observability condition in equation~\eqref{eq_formulation1} to structural observability. This relaxation is valid since the set of values for which structural observability does not match with observability lies on algebraic subspace with zero Lebesgue measure \cite{woude:03}. Such relaxation has been used in related literature \cite{asilomar11,liu-pnas,pequito2017structurally,pequito2013optimal,pequito2014optimal,pequito2015distributed,pequito_gsip, jstsp}. Related graph-theoretic concepts for structural observability are defined in the followings.

In Problem Formulation~\ref{prob_1}, consider $\mc{A} \sim \{0,1\}^{n \times n}$ as the structured  representation of system matrix $A$.
Nonzero elements of $\mc{A}$ are defined by system parameters, and the zeros are system fixed zeros. Similarly, $\mc{H} \sim \{0,1\}^{m \times n}$ is the structure of measurement matrix $H$.  $\mc{H}_{ij}$ being nonzero implies measurement of state $x_j$ by sensor $y_i$. In structured systems theory, such zero-nonzero structure of $\mc{A}$ and $\mc{H}$ is represented by a directed graph (digraph)  $\mc{G}_{sys} \sim (\mc{X} \cup \mc{Y},\mc{E})$, which is referred to as system digraph.  Note that, $\mc{X}$ is the set of state nodes $\{{x}_1,\hdots {x}_n\}$ each representing a state, and $\mc{Y}$ is the output set $\{{y}_1, \hdots {y}_m\}$ representing the sensor measurements. The nonzero entry $\mc{A}_{ij}$  is represented by an edge ${x}_j \rightarrow {x}_i $. The nonzero entry $\mc{H}_{ij}$  is represented by an edge ${x}_j \rightarrow {y}_i $.
Define a path as sequence of non-repeated nodes connected by edges represented by $\xrightarrow{path}$. Denote by $\xrightarrow{path} \mc{Y}$ a path that ends in an output node in $\mc{Y}$. A path that starts and ends at the same node is called a cycle.

A graph is called Strongly Connected (SC) if there is a path from every state $x_i$ to every other state $x_j$. If the graph is not SC it can be decomposed to Strongly Connected Components (SCC). Note that states in a SCC are mutually reachable, i.e. there is a path from every state to every other state in that SCC.
In order to explore states necessary for observability, we partition all SCCs in terms of their reachability by states in other SCCs. In this direction, SCC that has no outgoing edges to other SCCs is called \textit{parent SCC}. In other words, for $SCC_l$ if for all $x_i \in SCC_l$ there is no $x_j \notin SCC_l$ such that $x_i \rightarrow x_j$, then $SCC_l$ is a parent SCC. A non-parent SCC is called a \textit{child SCC}. In other words, if $SCC_l$ is a child SCC there exist $x_i \in SCC_l$ and $x_j \notin SCC_l$ such that $x_i \rightarrow x_j$.
Using this classification we state the main theorem on structural observability of structurally cyclic system\footnote{A system is structurally cyclic if there is a family of cycles spanning all state nodes. For such system the system matrix $A$ is full-rank.}.

\begin{theorem}  \label{thm_SCCp}
	Given a structurally cyclic system, the system is structurally observable if and only if  one state in every parent SCC is measured by a sensor, i.e. for every parent $SCC_l$ there exist $x_i \in SCC_l$ such that $x_i \rightarrow \mc{Y}$.
\end{theorem}

See detailed proof in our previous work \cite{asilomar11,jstsp}.
	
This theorem further implies that  the number of necessary sensors for observability equals the number of parent SCCs in structurally cyclic systems. In this direction, assigning a sensor measurement to every parent SCC satisfies structural observability.
Parent SCCs do not share any state node.
\textit{Note that Parent SCCs (in general all SCCs) do not share any state node.}
, and there are polynomial time algorithms to decompose the system digraph  into disjoint SCCs and define their type (parent or child), namely Depth First Search (DFS) algorithm \cite{algorithm} with computational complexity of $\mc{O}(m^2)$.

Next, we extend the structural observability to distributed case. Note that to recover distributed observability at every sensor the network $\mc{G}_W$ must be designed such that the system is observable to each sensor. Notice that the proper network design along with stochasticity of network adjacency matrix is sufficient for distributed observability. The sufficient condition on the network design is stated in the following theorem:

\begin{theorem} \label{thm_dist}
	For the system digraph, $\mc{G}_{sys}$ let the sensors have the necessary measurements based on Theorem~\ref{thm_SCCp}, i.e. every sensor has a measurement of a (distinct) parent SCC. The system is distributed observable in structural sense if for every sensor there is a directed path in the network to every other sensor, i.e. the sensor network $\mc{G}_W$ is SC.
\end{theorem}

The proof follows from the output-connectivity condition of structural observability. Note that for observability there must be a directed path from every parent SCC to every sensor. Since (i) there is a link from a state in every parent SCC to a distinct sensor node, i.e. for $SCC_l \rightarrow y_i$, and also there is a path from every sensor node to every other sensor node, i.e. $y_i \xrightarrow{path} y_j $ for $j \in \{1, ...,m\}$. Therefore for every $SCC_l,~l \in \{1, ...,m\}$ there is a path to every sensor node $y_i, ~i \in \{1, ...,m\}$. See detailed proof in \cite{jstsp,asilomar11,nuno-suff.ness}.

Notice that since all the costs $c_{ij}$ and $b_{ij}$ are positive, the minimization of $\sum_{i=1}^{m} \sum_{j=1}^{n} c_{ij}\mc{H}_{ij} + \sum_{i=1}^{m} \sum_{j=1}^{m} b_{ij}\mc{W}_{ij}$ can be separarted into minimization of $\sum_{i=1}^{m} \sum_{j=1}^{n} c_{ij}\mc{H}_{ij}$ and minimization of $\sum_{i=1}^{m} \sum_{j=1}^{m} b_{ij}\mc{W}_{ij}$. Second, based on Theorem~\ref{thm_SCCp} and~\ref{thm_dist}, the distributed observability constraint is equivalent to (i) having one sensor measurement from each parent SCC for structural observability of $(A,H)$ (Theorem~\ref{thm_SCCp}), and (ii) having the network of these sensors to be SC (Theorem~\ref{thm_dist}). Note that, the first constraint (i) is related to sensing cost optimization $\sum_{i=1}^{m} \sum_{j=1}^{n} c_{ij}\mc{H}_{ij}$ while the second constraint (ii) is related to networking cost optimization $\sum_{i=1}^{m} \sum_{j=1}^{m} b_{ij}\mc{W}_{ij}$. Therefere both optimization and constraint can be decomposed into separate cost optimization problems, which results in an exactly equivalent formulation as follows:
\begin{prob} \label{prob_2}
	
	\textbf{$\mc{P}1$}. For a group of sensors with $c$ as the measurement cost matrix, the problem is to minimize sensing  cost of the dynamical system \eqref{eq_sys1}:
	\begin{equation} \label{eq_formulation2_1}
	\begin{aligned}
	\displaystyle
	\argmin
	\limits_{\mc{H}} ~~ & \sum_{i=1}^{m} \sum_{j=1}^{n} c_{ij}\mc{H}_{ij} \\
	\text{s.t.} ~~ & (A,H)\mbox{-structural~observability},\\
	~~ &  \mc{H}_{ij} \in \{0,1\}\\
	~~ &  A~\mbox{is structurally full rank}\\
	\end{aligned}
	\end{equation}
	
	\textbf{$\mc{P}2$}. For the network of sensors with $b$ as the link cost matrix of $\mc{G}_{net}$, the problem is to minimize the networking cost as:
	\begin{equation} \label{eq_formulation2_2}
	\begin{aligned}
	\displaystyle
	\argmin
	\limits_{\mc{W}} ~~ & \sum_{i=1}^{m} \sum_{j=1}^{m} b_{ij}\mc{W}_{ij} \\
	\text{s.t.} ~~ & \mc{G}_W \subset \mc{G}_{net},~~\mc{G}_W~\mbox{is~SC} ~~~~~~~~~~\\
	~~ &  \mc{W}_{ij} \in \{0,1\}\\
	\end{aligned}
	\end{equation}
\end{prob}

\section{Combinatorial Optimization Solution} \label{sec_sol}
\subsection{Solution to Problem $\mc{P}1$ }
In this subsection, we solve the sensing cost optimization $\mc{P}1$ (for structurally full-rank systems). This problem is claimed to be NP-hard for general systems in \cite{pequito_gsip}. However, here we provide a polynomial order solution for $\mc{P}1$ for structurally cyclic systems.
To minimize the sensing cost for estimation, the assumption is that the number of sensors equals the number of necessary
measurements for structural observability. The minimal number of sensors for observability is primarily addressed in our previous work \cite{jstsp14, globalsip14,icassp2016}.

As mentioned in Section~\ref{sec-str} and following the stated assumptions,  the minimum number of sensors for structurally full-rank systems equals the number of parent SCCs in the system digraph, resulting the following formulation:

\begin{prob} \label{prob_3}
	
	\textbf{$\mc{P}1$}. Considering minimum number of sensor measurements for observability, the sensing cost optimization problem takes the following form:
	\begin{equation} \label{eq_formulation3}
	\begin{aligned}
	\displaystyle
	\argmin
	\limits_{\mc{H}} ~~ & \sum_{i=1}^{m} \sum_{j=1}^{n} c_{ij}\mc{H}_{ij} \\
	\text{s.t.} ~~ & (A,H)\mbox{-structural~observability},\\
		~~ &  \sum_{i=1}^{m} \mathcal{H}_{ij} \leq 1, ~~ \sum_{j=1}^{n} \mathcal{H}_{ij} = 1\\
		~~ &  \mc{H}_{ij} \in \{0,1\}\\
	\end{aligned}
	\end{equation}
\end{prob}
The extra conditions do not change the optimization problem. The added constraint $ \sum_{i=1}^{m} \mathcal{H}_{ij} \leq 1$ implies that all states are measured by at most one sensor, and $\sum_{j=1}^{n} \mathcal{H}_{ij} = 1$ follows the assumption that every sensor takes one state measurement. Next, following the  fact that parent SCCs are separate components, the problem can be reformulated as assigning a group of sensors to take measurement of a group of parent SCCs.
For this formulation, let define a new cost matrix $\mathcal{C}_{m \times m}$, where  $\mathcal{C}_{ij}$ denotes the cost of assigning a parent, $SCC_j$, to sensor $y_i$. This cost is defined as the \textit{minimum} sensing cost among all  states in parent $SCC_j$, i.e. $\mathcal{C}_{ij}= \min \{c_{il}\},~ x_l \in  SCC_j,~ i,j \in \{1, \hdots, m\}$
The above equation reformulates matrix $c_{m \times n}$ to matrix $\mathcal{C}_{m \times m}$. In other words,  the cost matrix relating sensors to states is transferred to cost matrix relating sensors to parent SCCs. In this direction, we introduce a new structured matrix $\mathcal{Z}\sim \{0,1\}^{m \times m}$. This matrix defines the assignment of sensors to parent SCCs, i.e. $\mathcal{Z}_{ij} \neq 0$ implies  a state in $SCC_j$ is measured by sensor $i$ ($ SCC_j \rightarrow y_i $). By recalling Theorem~\ref{thm_SCCp}, observability is guaranteed by sensing all parent SCCs. Hence, following setup represents the new formulation of original problem $\mc{P}1$:
 \begin{prob} \label{prob_final}
 	\textbf{$\mc{P}1$}. let have $m$ sensors to be assigned to $m$ parent SCCs in a  structurally cyclic systems; the sensing cost optimization is reformulated as:
 	\begin{equation}
 	\begin{aligned}
 	\argmin
 	 \limits_{\mathcal{Z}} ~~  & \sum_{i=1}^{m} \sum_{j=1}^{m} (\mathcal{C}_{ij}\mathcal{Z}_{ij}) \\
 	\text{s.t.} ~~ &  \sum_{j=1}^{m} \mathcal{Z}_{ij} = 1, ~~~ \sum_{i=1}^{m} \mathcal{Z}_{ij} = 1\\
 	~~ &  \mathcal{Z}_{ij} \in \{0,1\} \\
 	\end{aligned}
 	\label{minlsap}
 	\end{equation}
 \end{prob}
 The new constraint $ \sum_{j=1}^{m} \mathcal{Z}_{ij} = 1$ means that every parent SCC is measured, which guarantees observability. Note that, formulation  \eqref{minlsap} represents a famous combinatorial optimization problem, known as Linear Sum Assignment Problem (LSAP) \cite{assignmentSurvey}. The most recent and most computationally efficient solution to the LSAP is known as \textit{Hungarian method} proposed in \cite{edmondsHungarian} with the complexity order of $\mathcal{O}(m^3)$. Therefore, noting that  Problem Formulation~\ref{prob_final} represents the relaxation to original problem $\mc{P}1$ in Problem Formulation~\ref{prob_2}, the sensing cost optimization $\mc{P}1$ finds a polynomial order solution. Note that structural relaxations hold for almost all numerical values of system parameters \cite{woude:03}. In other words, for any choice of system/measurement/network matrices as long as the structures are fixed (with potentially time-varying entries) and structural results consists with the LSAP, the relaxed solution is exact and the gap is zero.
 
 \subsection{Solution to Problem $\mc{P}2$ }
In this subsection, we discuss the solution for networking cost optimization problem $\mc{P}2$ stated  by equation \eqref{eq_formulation2_2}. Recall that in this problem the goal is to find the minimum cost Strongly-Connected (SC) subgraph spanning network $\mc{G}_{net}$.
We separately discuss the solution of $\mc{P}2$ for (i)~undirected networks, and (ii)~directed networks.

\textbf{Undirected network $\mc{G}_{net}$:}
Consider the case that every communication link in network of sensors is bidirectional and the networking cost matrix $b$ is symmetric. Note that in this case we assume the sensors share their information mutually, i.e. sensor $i$ shares its prediction with sensor $j$ if and only if sensor $j$ shares its prediction with $i$.
In this case,  $\mc{P}2$ represents a known problem in combinatorial optimization and is equivalent to the  Minimum  Spanning Tree (MST) or Minimum Weight Spanning Tree problem \cite{gabow1986efficient}. This problem is known to have polynomial order solution using \textit{Prim's algorithm} \cite{prim1957shortest} (or \textit{Kruskal's algorithm} \cite{kruskal1956shortest}). Given the adjacency matrix of $\mc{G}_{net}$ the computational complexity of the Prim's algorithm is $\mc{O}(m^2)$. Note that, similar to the solution to $\mc{P}1$, in this case the structural relaxed solution based on Kruskal's or Prim's algorithm is exact and the gap is zero. 

\textbf{Directed network $\mc{G}_{net}$:}
Consider the case that the network links are directed  and   $\mc{G}_{net}$ represents a directed graph (digraph). In this case,  $\mc{P}2$ represents the Minimum Spanning Strong Sub(di)graph (MSSS) problem, which is known to be NP-hard \cite{digraphs}\footnote{It is known that the MSSS problem generalizes the Hamiltonian-Cycle problem and therefore is NP-hard \cite{digraphs}.}. Approximations to solve this problem are proposed in the literature. In \cite{frederickson1981approximation}, the authors proposed a polynomial-time algorithm as follows: fix a root node of the directed network and then find the \textit{in-branching} (also known as \textit{arborescence}) and \textit{out-branching}\footnote{The in-branching is a generalization of MST to digraphs where each node has a minimum cost directed path to the root node. The out-branching is the reverse problem where the root node has minimum cost directed path to every other node \cite{digraphs}.} and take the union of the two branching as the MSSS. Note that the cost   of each branching in worst case equals to MSSS. Therefore the output of this approach gives an approximation at factor of $2$. \textit{Edmond's algorithm} \cite{edmonds1967optimum} is the efficient way to find the in/out-branching with computational complexity of $\mc{O}(m^2)$. Let define gap as follows: if $\mc{F}$ is the outcome of the relaxed algorithm in worst-case (i.e. the upper-bound for the solution) and $\mc{L}$ is the exact minimal solution,  the gap is defined as: $\frac{\mc{F}-\mc{L}}{\mc{L}}$. Then the gap of the Edmond's algorithm is $1$.

\section{Conclusion}

In this paper, we proved a polynomial order solution of  complexity $\mc{O}(m^3)$ for sensing cost optimization in structurally full-rank systems. A $2$-approximation polynomial order solution (with gap $1$) with $\mc{O}(m^2)$ complexity  is provided for generally NP-hard networking cost optimization. Further, considering undirected network of sensors an exact polynomial order $\mc{O}(m^2)$ solution is proposed for this problem. 

\bibliographystyle{IEEEbib}
\bibliography{bibliography}
\end{document}